\documentstyle[12pt]{article}
\def\GA{\raise2.5pt\hbox{$>$}\kern-8pt\lower2.5pt\hbox{$\sim$}}
\def\LA{\raise2.5pt\hbox{$<$}\kern-8pt\lower2.5pt\hbox{$\sim$}}
\def\LI{${\sf L}_{\sf 1}$}
\def\LII{${\sf L}_{\sf 2}$}
\def\LIII{${\sf L}_{\sf 3}$}
\def\BF{{\sf BF}}

\title{Disordered
Regimes of the one-dimensional
complex Ginzburg-Landau equation
\vspace{48pt}
}

\author{Hugues Chat\'e\\ \\
\normalsize
{\it Service de Physique de l'Etat Condens\'e}\\
\normalsize
{\it Centre d'Etudes de Saclay}\\
\normalsize
{\it 91191 Gif-sur-Yvette, France.}}

\date{\mbox{ }}

\vspace{24pt}

\begin{document}
\maketitle

\begin{abstract}
\noindent
I review recent work on the ``phase diagram'' of the one-dimensional complex
Ginzburg-Landau equation for system sizes at which chaos is extensive.
Particular attention is paid to a detailed description of the spatiotemporally
disordered regimes encountered. The nature of the transition lines separating
these phases is discussed, and preliminary results are presented which aim
at evaluating the phase diagram in the infinite-size, infinite-time,
thermodynamic limit.
\baselineskip 18pt

\end{abstract}

\vfill
{\sl To appear in ``Spatiotemporal Patterns in Nonequilibrium
Complex Systems'', Santa Fe Institute Series in the Sciences of Complexity,
Addison-Wesley, 1993.}
\thispagestyle{empty}

\newpage

\vspace{24pt}
\noindent
{\large \bf 1. A prototype for studying spatiotemporal chaos}
\vspace{12pt}

\noindent
The complex Ginzburg-Landau equation (CGL):
$$
\partial_t A = A + (1+ib_1) \Delta A - (b_3-i) |A|^2 A
$$
where  $A$ is a complex field, $x\in[0,L]$ and $b_3>0$,
is of considerable importance
to everybody interested in spatially extended non-equilibrium
systems$^5$.
It  accounts for the slow modulations, in space and time, of the oscillatory
state in a physical system which has undergone a Hopf bifurcation$^{10}$.
As such, the CGL is closely  related to numerous experimental
situations$^5$.
The resulting universality and genericity are
accompanied by specific features
which make this model interesting for its own sake, simply as a prototype
of spatially extended dynamical systems.
\baselineskip 18pt

There exist
two important limits to CGL: the dissipative, relaxational
``real Ginzburg-Landau''
equation ($b_1=0, b_3 \rightarrow \infty$) and the dispersive,
integrable nonlinear
Schr\"odinger equation (NLS) ($b_3=0, b_1 \rightarrow \infty$).
Studying CGL in a comprehensive manner, one encounters dynamical
regimes where the relative importance of dissipation and dispersion
can be tuned at will in a non-integrable, non-variational system.
This has been done recently
for the one-dimensional case, uncovering the ``phase diagram'' presented in
figure~1 and discussed below$^{12,2}$.

The relative simplicity of figure~1
stems mostly from the existence
of a ``thermodynamic limit'' for the spatiotemporally chaotic regimes observed.
As a matter of fact, away from the intricacy of the bifurcation diagrams
at small sizes ($L\LA 50$), there exists a large-size limit beyond which
chaos becomes extensive and can be characterized by intensive quantities
independent of system size, boundary conditions, and, to a large extent,
initial conditions. Perhaps the most convincing illustration, for CGL, of this
essential property of spatiotemporal chaos can be found in a recent paper by
Egolf and Greenside$^6$
where they show that the Lyapunov dimension
is proportional to the system size $L$.

Here, I review the various disordered ``phases''
observed in the one-dimen-sional
case in the large-size limit and
the nature of the transitions leading to them. Emphasis is put
on the physical-space structures and objects which compose the spatiotemporal
chaos as they could well play a crucial role when trying to build
a statistical analysis of the disordered phases. This should also
hopefully provide a clearer picture of the ``elementary mechanics''
at play in these regimes.

All the results reported and presented here were obtained using
a pseudo-spectral code with periodic boundary conditions and second-order
accuracy in space and time. Spatial resolution was typically 1024 modes
for a domain of size $L=1000$. Time steps varied like $1/\sqrt{b_3}$,
with typical values as indicated in the figure captions.

\vspace{24pt}
\noindent
{\large \bf 2. Phases}
\vspace{12pt}

\noindent
Early work on CGL
has dealt with the problem of the linear stability
of its family of plane-wave solutions
$A=a_k \exp i(kx+\omega_k t)$
with $a_{k}^2=(1-k^2)/b_3$ and $\omega_k=1/b_3-(b_1+1/b_3)k^2$.
All these solutions are unstable for $b_1>b_3$, a condition which defines
the so-called ``Benjamin-Feir line'' ({\sf BF}).
For $b_1<b_3$, plane-wave solutions with
$k^2<k_{\rm Eckhaus}^2 = (b_3-b_1)/(3b_3-b_1+2/b_3)$ are linearly
stable$^8$.
The {\sf BF} line thus provides a first separation of the $(b_1,b_3)$
plane, indicating where the constant-modulus, phase-winding solutions
might play a role.

I now give a brief account of what is known about the disordered phases.
For all these spatiotemporal chaos regimes, chaos is extensive: the Lyapunov
dimension (and the other quantities derived from the Lyapunov spectrum such
as the entropy or the number of positive exponents) scales linearly with
the system size $L$.$^6$

\vspace{12pt}
\noindent
{\sl \bf 2.1 Defect turbulence}
\vspace{12pt}

\noindent
Above the {\sf BF} line and to the left of \LI, a strongly disordered
phase is observed, best defined by a finite density of space-time zeros
of $|A|$. This ``defect turbulence'', also named ``amplitude
turbulence'', is characterized by a quasi-exponential decay of  space and time
correlation functions, and  slower-than-exponential tails of the
pdf of the phase gradient $\phi_x$.$^{12}$

One elementary process is the key ingredient of this strong spatiotemporal
chaos: pulses of $|A|$ grow under the effect of  the dispersion term;
this ``self-focusing'' is stopped
by the action of dissipation, breaking the pulse. Such events can be seen
in the spatiotemporal plots presented in figure~2; regions of almost
zero amplitude are present, another indication that pulses are the relevant
objects to consider when approaching the NLS limit.
As one gets
closer to the NLS (increasing $b_1$), the pulses get sharper and higher,
whereas
closer to the \BF~ line, dissipation dominates and $|A|$ rarely overpasses
the ``saturation'' value $1/\sqrt{b_3}$.

This disordered phase is reached from (almost) all initial conditions
for parameter values to the left of \LIII. In the triangular region
delimited by \BF, \LI~ and \LIII. it coexists with ``phase turbulence'',
which is described below.

\vspace{12pt}
\noindent
{\sl \bf 2.2 Phase turbulence}
\vspace{12pt}

\noindent
Phase turbulence is a weakly disordered regime observed in the region of
parameter space above the \BF~ line to the right of \LI, as well as in the
triangular region delimited by \BF, \LI~ and \LIII. It is best defined
by the absence of space-time defects or, equivalently, by the
bounded character of the pdf of $\phi_x$ and $|A|$ (see below for a
 discussion of this point). The absence of
phase singularities implies that the ``winding number''
$\nu=\int_{0}^{L} \phi_x dx$ is a conserved quantity equal to a multiple of
$2\pi$ which classifies the different attractors reached at given values
of $b_1$ and $b_3$.

Chaos is very weak, as shown by the slow decay of
space and time correlation functions, indicative of diffusive or
sub-diffusive modes.$^{12}$
This is corroborated by the flat shoulder of the
spatial power spectrum of $|A|$ at low wavenumbers, reminiscent of the
Kuramoto-Sivashinsky equation (KS).
The dynamics is in fact very similar to that of KS, which
is not surprising since this equation was originally derived to describe
the phase dynamics of CGL near the \BF~ line.$^{10}$
The close relationship with KS suggests, in turn, that the Kardar-Parisi-Zhang
interface equation (KPZ)$^9$, which have been argued to account for
the large-scale,
long-time, properties of KS,$^{13}$ might provide the correct description
of the long-wavelength limit of phase turbulence. In such a setting,
one expects exponential decay of spatial correlations and stretched-exponential
decay of temporal correlations.$^9$

Spatiotemporal diagrams (figure~3)
reveal that the objects involved are ``shocks''
of $|A|$ (similar graphs for $\phi_x$ show localized modulations
of the wavenumber, and look essentially the same). Close to the \BF~ line,
the dynamics is indeed very much like in KS (figure~3a),
while away from it trains of
propagative shocks become more and more frequent (figure~3b).
The overall effect of $\nu$ is to introduce a general drift on the localized
shocks of $|A|$ which increases with ${\rm abs}(\nu)$ (figure~3c).
The phase turbulence regimes thus look very different depending on $\nu$
and the distance, in the parameter plane, from $(b_1,b_3)$ to the \BF~ line.
Increasing this distance, the characteristic scales  decrease
(the difference of the time-scales between figures~3a and 3b is only partly
due to the variation of the basic frequency like $\sqrt{b_3}$).
The other important effect, the appearance of trains of propagative shocks,
is an indication that KS is {\it not} a valid description away from the
\BF~ line and that other terms, incorporating odd derivatives of $\phi$,
are present in the ``effective phase equation'' describing the corresponding
phase turbulence regimes. Whether the connection to KPZ still holds in
such circumstances and what is then the valid phase equation replacing KS,
are open problems under investigation.

It must be mentioned, at this point, that it is still a controversial and
unresolved issue to know whether phase turbulence ``exists'' in the
thermodynamic limit, i.e., whether it is a transient phenomenon or a
finite-size artifact. In sections~3.1 and 3.3 below, I discuss this point
in relationship with the nature of the transition lines delimiting the
phase turbulence region in figure~1. In any case, even if there always
exists a very small (essentially undetectable) density of defects, the
spacetime regions between those rare events are large enough so as to justify
the study of phase turbulence for its own sake.

\vspace{12pt}
\noindent
{\sl \bf 2.3 Spatiotemporal intermittency}
\vspace{12pt}

\noindent
Below the \BF~ line, plane-wave solutions of wavenumber
$k^2 \le k^2_{\rm Eckhaus}$ are linearly stable. This  does not preclude
the existence, to the left of \LII, of other,  mostly chaotic, solutions
which are easily reached for initial conditions outside the basin of attraction
of the stable plane waves.$^2$ These disordered regimes
are {\it spatiotemporal intermittency} regimes:
they consist of space-time regions
of stable plane  waves separated by localized objects evolving and interacting
in a complex manner (figure~4).
The plane waves constitute the passive, ``absorbing''
state while the localized objects carry the spatiotemporal disorder. The
simplest
of these objects are members of the family of exact solutions found by
Nozaki and Bekki.$^1$
These propagating ``holes'' of $|A|$ are not defects, strictly
speaking, even though $|A|$ might remain very small in
their core.$^2$
Defects (zeros of $A$) do occur, but at scattered points in spacetime.
This makes the density of defects difficult to estimate and
rather worthless, since, for example, the average number (in space)
of hole-like
objects is probably a better characterisation of the dynamics. Likewise,
space and time correlation functions are not the easiest quantities
to measure when evaluating the coherence scales in these regimes, due to
the intermittent character of the disorder.
As usual with spatiotemporal intermittency,$^4$ one can take advantage
of the intrinsic binary structure of the spatiotemporal dynamics and evaluate,
say, the distributions of sizes and lifetimes of the patches of plane wave
solutions. These distributions are roughly exponential, yielding
characteristic coherence scales.$^2$

The spatiotemporal intermittency regimes do exhibit space-time defects. To that
extent, they may be considered as defect turbulence regimes. The key difference
with the region of defect turbulence described above is that here
the defects do not appear spontaneously, they are produced by the localized
objects carrying the disorder.

\vspace{12pt}
\noindent
{\sl \bf 2.4 The ``bichaos'' region}
\vspace{12pt}

\noindent
The domain of parameter space delimited by the \BF, \LI~ and \LIII~ lines
deserves further comment. Depending on the initial conditions,
one can reach one of two spatiotemporally chaotic regimes, phase
turbulence or defect turbulence. A closer look at the defect turbulence
reveals that it consists of localized propagating and branching
hole-like objects which separate more quiescent space-time regions (figure~5).
These objects do not appear spontaneously. The
quiescent regions are nothing else than patches of phase turbulence, with
their characteristic shocks of $|A|$. Defect turbulence  is thus
 a spatiotemporal intermittency regime, with the absorbing state
being the phase turbulence regime.

\vspace{24pt}
\noindent
{\large \bf 3. Transitions}
\vspace{12pt}

\noindent
I now examine the various transition lines between the phases and discuss
to what extent they can be considered as phase transitions.

\vspace{12pt}
\noindent
{\sl \bf 3.1 The \LI~ line}
\vspace{12pt}

\noindent
Line \LI~ separates defect turbulence from phase turbulence for values
of $b_1 \GA 1.85$. As such, its best definition is given by the parameter
values for which the density of defects $n_D$ goes to zero.
It was shown$^{12}$ that, for $b_1=3.5$, $n_D \sim (b_3-b_3^*)^2$ with
$b_3^*\simeq 1.29$. No hysteresis has been detected. This transition thus
appears continuous, even though one cannot exclude a crossover scenario
where $n_D$ would remain small beyond $b_3^*$.

The situation is not as clear from the point of view of the correlation length
$\xi$ deducted from the exponential decay of the two-point correlation
function: approaching \LI, $\xi$ increases but does not diverge
at the transition point.$^{12,7}$
Although difficult to measure in the phase turbulence
regimes, indirect arguments imply that $\xi$ remains finite.
Only a change of behavior is observed,
apparently accompanied by a discontinuity of ${\rm d}\xi/{\rm d}b_3$.
A similar situation holds for the variation of the Lyapunov dimension
and the other quantities derived from the Lyapunov spectrum when crossing
\LI, as shown by Egolf and Greenside.$^6$
In other words, no critical behavior seems to be present at the transition,
even though it is continuous.

This remark is valid for the experimental conditions within which the above
results were obtained. In particular, the system sizes used ensure
the extensivity of chaos, but no assessment of finite-size effects has been
made in order to evaluate the infinite-size, infinite-time, ``thermodynamic''
limit. The main question in this respect is that of the position of \LI~
as $L\rightarrow\infty$. If \LI~ remains away from the \BF~ line, the existence
of phase turbulence in the thermodynamic limit is assured and the
 transition at \LI~ does not show any critical behavior. Another hypothesis
consists in  \LI~ moving toward \BF~ as $L$ increases, with the correlation
length at threshold increasing and finally diverging for an infinite-size
system. In this case, the phase turbulence region disappears in the
thermodynamic limit and the \BF~ line is the asymptotic boundary of
defect turbulence.

Preliminary results (figure~6) indicate a clear displacement of \LI~
toward \BF~ as $L$ increases, making the second hypothesis more likely.
A similar trend was recently found by Egolf and Greenside.$^7$
Extensive numerical work is under completion to obtain a quantitative
extrapolation of the position of \LI~ in the infinite-size limit.

\vspace{12pt}
\noindent
{\sl \bf 3.2 The \LII~ line}
\vspace{12pt}

\noindent
The line \LII, best defined as the limit of existence
of spatiotemporal disordered states, is difficult to determine in a precise
manner.$^2$
This is mostly due to the appearance, in the transition region,
of ``frozen'' states, i.e. spatially-disordered arrangements of localized
objects with trivial time dependence (figure~7). The coherence scales of the
spatiotemporal intermittency regimes increase when approaching \LII,
but their divergence is not observed; the system ``falls'' on a frozen state.
As often in spatiotemporal intermittency situations$^3$,
the deterministic
features of the system ``mask'' the directed-percolation-like phase
transition. The line \LII~ is thus determined only crudely, awaiting
further progress.
In particular, two directions seem worth investigating:
a ``local'' approach based on the study of the interactions
between the localized objects at the origin of the
spatiotemporal disorder, and a ``global'' approach based on Lyapunov analysis
of the disordered regimes.

\vspace{12pt}
\noindent
{\sl \bf 3.3 The \LIII~ line}
\vspace{12pt}

\noindent
The \LIII~ line delimits the parameter space region where phase turbulence
ceases to coexist with defect turbulence. Here I give a preliminary account of
ongoing work on the nature of the breakdown of phase turbulence occurring
when crossing \LIII~ (say, by decreasing $b_3$), for a winding number
$\nu=0$.

The transition is hysteretic: defect turbulence persists when crossing \LIII~
back, up to line \LII. The breakdown of phase turbulence$^{11}$
occurs {\it via} the nucleation of a first defect, followed by
the quasi-deterministic invasion of the phase turbulent state by the defect
phase (figure~8).
The speed of invasion is always finite and rather large; it is given
by the average propagation speed $v_D$ of the localized objects composing the
defect phase in this region of parameter space. The precursor of the
first defect is apparently a local event: one of the shocks of $|A|$
composing the phase turbulence accelerates, and a depression of $|A|$
develops at its tail. When the velocity reaches $v_D$, the first hole-like
object typical of the defect phase is effectively created, and quickly
generates the first space-time defect.
It is difficult to determine what triggers the initial acceleration
of one of the shocks of $|A|$. As mentioned above, more and more trains
of propagating shocks are observed in the phase turbulence regimes as
one goes away, in parameter space, from the \BF~ line.
Simultaneously,
the average velocity $v_S$ of these propagative shocks increases.
A possible interpretation
of the nucleation is that, for parameter values beyond \LIII,
the fluctuations around $v_S$ bring the speed of one shock past
a critical value, after
which the shock is attracted to the hole-like solution characteristic of the
defect phase. Such a critical value would be related to correponding
critical values of the local minimum of $|A|$ and local extremum
of $\phi_x$, since as the velocity of the shock increases, the local
depression of $|A|$ at its tail deepens, and the local phase gradient
increases.
Ongoing work aims at making those mostly qualitative statements
more quantitative.

As described above, the breakdown of phase turbulence strongly resembles
a first-order phase transition. So far, \LIII~ has been determined numerically
by merely checking whether the breakdown occurs within a fixed (very long)
integration time.$^{12}$ However crude, this methodology produced a sharp
transition line, all the more so as no size effects could be detected.

A quantitative estimate of the probability  of nucleation is needed to improve
this situation. Preliminary results on the study of the pdfs of $|A|$ and
$\phi_x$ in the phase turbulence regimes provide the beginning of an answer
(figure~9).
Away from \LIII~  (figures~9a and 9c), these distributions
have either Gaussian or strictly bounded tails.
Close to \LIII~  (figures~9b and 9d), these tails are exponential,
signalling a clear change of behavior
on which the definition of \LIII~ can be firmly based.
Nevertheless, the question of the Gaussian/bounded character of the tails
away from \LIII~ remains a crucial one since it determines whether the
probability of nucleation can be strictly zero or not.
A plausible scenario would consist of exponential  tails truncated
by a finite bound moving to infinity (resp. zero) for $\phi_x$ (resp. $|A|$)
when approaching \LIII. Whereas this scenario would assure the existence
of the phase turbulence in the infinite-size, infinite-time limit, the other
scenario of a crossover from Gaussian to exponential tails would not.
Current numerical efforts aim at providing data to decide this.

In the same spirit,
another important question deals with the estimation of size-effects in this
region of parameter space.
So far, results for sizes $L=1000$ and $L=4000$ have not shown any
significant difference, strengthening the case for a first-order transition.

Finally, the overall effect of a non-zero winding number $\nu$ is to give
rise to earlier breakdown of the phase turbulence regime, moving
\LIII~ to the right, as expected from the general additional drift
of the shocks in this case.

\vspace{24pt}
\noindent
{\large \bf 4. Open problems}
\vspace{12pt}

\noindent
The ``phase diagram'' presented in figure~1 was obtained for systems sizes
$L$ large (for which the extensivity of chaos is ensured) but finite.
{}From the above discussion of lines \LI~ and \LIII, it appears that an
estimation
of the finite-size effects is necessary, and even crucial, since the
existence of phase turbulence depends on the position and nature of these
transition lines in the infinite-size, infinite-time, ``thermodynamic'' limit.
For line \LIII, a detailed study of the pdfs of interest in the phase
turbulence regimes should be able to provide an objective criterion
which can help to determine precisely
the breakdown of phase turbulence.
In this respect, preliminary results (figure~9) are encouraging.
For the line \LII~ delimiting the domain of existence of spatiotemporal
intermittency, the key point lies in finding a quantifier which can be
used to determine its position efficiently. Quantities related to the Lyapunov
spectrum might constitute such a quantifier.

Apart from these ``global'', statistical quantities,
approaches based on the study of the local objects involved in the
spatiotemporally disordered dynamics (``shocks'' for phase turbulence,
``pulses'' and ``defects'' for defect turbulence, ``sources'' and ``sinks''
for spatiotemporal intermittency) might also help to clarify some of
the problems left open.

At any rate, and whatever the situation is in the thermodynamic limit,
the ``finite-size'' phase diagram of figure~1 is of interest to all
``experimentalists'' (i.e., anybody working with finite-size systems
observable during a finite time). In this respect, it shows how varied and
complex statiotemporally chaotic regimes can emerge from such a simple
equation as CGL. I hope the above review will also provide guidelines for
dealing with similar cases, in particular when the question of what to measure
arises.

\vspace{18pt}
I thank D.A.~Egolf and H.S.~Greenside for stimulating exchanges and
the communication of their unpublished results, and P.~Manneville
for many discussions and comments.

\newpage
\vspace{24pt}
\noindent
{\large \bf References}
\vspace{12pt}

\begin{itemize}

\item[$^1$] Bekki, N., and K.~Nozaki, ``Exact Solutions of the Generalized
Ginzburg-Landau Equation.'' {\it J. Phys. Soc. Jap.}
 {\bf 53} (1984): 1581-1582;
``Formation of Spatial Patterns and Holes in the Generalized Ginzburg-Landau
Equation.'' {\it Phys. Lett. A} {\bf 110} (1985): 133-135.

\item[$^2$] Chat\'e, H., ``Spatiotemporal Intermittency Regimes
of the One-Dimensional Complex Ginzburg-Landau Equation'', to appear
in {\it Nonlinearity} 1993.

\item[$^3$] Chat\'e, H., and P.~Manneville, ``Role of Defects in the
Transition to Turbulence via Spatiotemporal Intermittency.''
{\it Physica D} {\bf 37} (1989): 33-41.

\item[$^4$] Chat\'e, H., and P.~Manneville
``Spatiotemporal Intermittency in Coupled Map Lattices.''
{\it Physica D} {\bf 32} (1988): 409-422;
Chat\'e, H., ``Subcritical Bifurcations and Spatiotemporal Intermittency.''
In: {\it Spontaneous Formation of Space-Time Structures and Criticality},
edited by T.~Riste and D.~Sherrington, 273-311. Dordrecht: Kluwer, 1991.

\item[$^5$] Cross, M.C., and Hohenberg, P.C., ``Pattern Formation
Outside of Equilibrium'', to appear in {\it Rev. Mod. Phys.} 1993,
and references therein.

\item[$^6$] Egolf, D.A. and H.S.~Greenside, ``Complexity
Versus Disorder for Spatiotemporal Chaos.'' Preprint available from the LANL
database ``chao-dyn@xyz.lanl.gov'' under number \# 9307010. See also
the abstract of a poster by the same authors in this volume.

\item[$^7$] Greenside, H.S., private communication.

\item[$^8$] Janiaud, B., A.~Pumir, D.~Bensimon, V.~Croquette,
H.~Richter, and L.~Kramer, ``The Eckhaus Instability for Travelling Waves.''
{\it Physica~D} {\bf 55} (1992): 259-269.

\item[$^9$] Kardar, M., G.~Parisi and Y.C.~Zhang, ``Dynamic Scaling of
Growing Interfaces''.
{\it Phys. Rev. Lett.} {\bf 56} (1986) 889-892.

\item[$^{10}$] Kuramoto, Y., {\it Chemical Oscillations,
Waves and Turbulence}. Tokyo: Springer, 1984;
Lega, J., ``D\'efauts Topologiques Associ\'es \`a la Brisure
de l'Invariance de Translation dans le Temps.'' Th\`ese de Doctorat,
Universit\'e de Nice, 1989; Coullet, P., L.~Gil, and J.Lega ``Defect
Mediated Turbulence.'' {\it Phys. Rev. Lett.} {\bf 62} (1989): 1619-1622.

\item[$^{11}$] Sakaguchi, H., ``Breakdown of the Phase Dynamics.''
{\it Prog. Theor. Phys.}  {\bf 84} (1990): 792-800.

\item[$^{12}$] Shraiman, B.I., A.~Pumir, W.~van~Saarloos, P.C.~Hohenberg,
H.~Chat\'e, and M.~Holen, ``Spatiotemporal Chaos in the One-Dimensional
Complex Ginzburg-Landau Equation.'' {\it Physica~D} {\bf 57} (1992): 241-248;
Pumir, A.,
B.I.~Shraiman, W.~van Saarloos, P.C.~Hohenberg, H.~Chat\'e, and M.~Holen,
``Phase vs. Defect Turbulence in the One-Dimensional Complex
Ginzburg-Landau Equation.'' In: {\it Ordered and Turbulent Patterns in
Taylor-Couette Flows}, edited by C.D.~Andereck.
New York: Plenum Press, 1992.

\item[$^{13}$] Yakhot, V., ``Large-Scale Properties of Unstable Systems
Governed by the Kuramoto-Sivashinsky Equation.'' {\it Phys. Rev. A.}
{\bf 24} (1981): 642-644;
Zaleski, S., ``A Stochastic Model for the Large Scale Dynamics of Some
Fluctuating Interfaces.'' {\it Physica D} {\bf 34} (1989): 427-438;
Procaccia, I., M.H.~Jensen, V.S.~L'vov, K.~Sneppen and R.~Zeitak,
``Surface Roughening and the Long-Wavelength Properties of the
Kuramoto-Sivashinsky Equation.'' {\it Phys. Rev. A.} {\bf 46} (1992):
3220-3224;
Sneppen, K., J.~Krug, M.H.~Jensen, C.~Jayaprakash and T.~Bohr,
``Dynamic Scaling and Crossover Analysis for the Kuramoto-Sivashinsky
Equation.'' {\it Phys. Rev. A} {\bf 46} (1992): 7351-7354.

\end{itemize}

\newpage
\noindent
{\large \bf Figure Captions}

\vspace{12pt}
\noindent
\begin{itemize}

\item[\bf Figure~1:] Phase diagram in the $(b_1,b_3)$ parameter plane.
Lines \LI~ and \LIII~ were determined in $^{12}$, line \LII~
in $^2$. See $^2$ for numerical details.

\item[\bf Figure~2:] Spatiotemporal representation of a defect turbulence
regime in a system of size $L=250$ for $b_1=3.5$ and $b_3=0.5$ (periodic
boundary conditions, timestep $\delta t =0.02$).
Time is running upward for $\Delta T=82.5$ (transient discarded).
Part (a) shows the evolution of $|A|$ (grey scale between $|A|=0$/black
and $|A|=2$/white). Part (b) shows the corresponding evolution of
$\phi=\arg(A)$
(grey scale between $0$/white and $\pi$/black).

\item[\bf Figure~3:] Spatiotemporal evolution of $|A|$ in phase turbulence
regimes in a system of size $L=1000$ for $b_1=3.5$
(periodic boundary conditions, timestep $\delta t=0.16$,
transient discarded, time running upward, grey
scale with white corresponding to maximum value of $|A|$).
Part (a): close to the \BF~ line ($\nu=0$, $b_3=2$,
$\Delta T=3300$, $0.68\le |A| \le 0.75$).
Part (b): close to \LI~ ($\nu=0$, $b_3=1.4$,
$\Delta T=1650$, $0.69\le |A| \le 1.01$).
Part (c): same as (a) but with $\nu=-10\pi$.

\item[\bf Figure~4:] Spatiotemporal intermittency regime in a system of size
$L=250$ (periodic boundary conditions, timestep $\delta t=0.04$)
for $b_1=0$ and $b_3=0.5$. Time is running upward for $\Delta T=165$
(transient discarded).
Same representation as in figure~2, but $0\le |A| \le 1.413$ in part~(a).

\item[\bf Figure~5:] Defect turbulence in the bichaos region ($L=1000$,
$b_1=1.5$, $b_3=1$, $\Delta T=1312$, periodic boundary conditions,
timestep $\delta t=0.04$, and
transient discarded). Spatiotemporal representation of $|A|$
in grey scale from $|A|=1.12$ (white) to $|A|=0$ (black); time is
running upward. Note that the hole-like propagating and branching objects
typical of the defect phase do not appear spontaneously in the phase
turbulent medium.

\item[\bf Figure~6:] Variation of the defect density $n_D$ ($\times 10^5$
on the graph) with $b_3$ near \LI. These results were obtained from runs of
duration $\Delta T\sim 20000$ after transients on systems of size $L=1000$ and
$2000$ with periodic boundary conditions and timestep $\delta t=0.04$.
Although these results are
still not very precise
(at such low densities, much longer runs would be necessary),
the general trend is significant: higher densities are reached for larger
sizes.

\item[\bf Figure~7:] Frozen state observed near \LII~ in a system of size
$L=500$ with $b_1=-0.9$ and $b_3=0.18$ (periodic boundary conditions).
This spatially disordered state is made of zero-velocity Nozaki-Bekki
holes and shocks. $|A|$ is stationary, while $\phi$ winds regularly
along time.

\item[\bf Figure~8:] Breakdown of phase turbulence past the \LIII~
 line in a system of size $L=1000$ for $b_1=1$, $b_3=0.675$ and $\nu=0$
(periodic boundary conditions, timestep $\delta t=0.16$).
Initial condition was a phase turbulent state for
$b_3=0.7$. The breakdown occurred near $t=12000$ as shown by the time series
of $|A|$ (part~(a)).
Part (b) is the spatiotemporal evolution of $|A|$ for
$t\in[9600,12800]$ (grey scale from $|A|=1.29$/white to $|A|=1.1$/black, time
running upward); at this resolution, the details of the defect phase are
not visible.
Part (c): zoom of (b) with a grey scale and a resolution
adapted to the defect phase ($x\in[500,1000]$, $t\in[12550, 12800]$,
$0\le |A| \le 1.29$); the nucleation of the first defect is obvious.
Part (d): as in (b) but for the locus of the minimum (in space) of $|A|$.

\item[\bf Figure~9:] Probability distribution functions of $|A|$ for phase
turbulence regimes in the bichaos region. These distributions were
obtained from the simulation of a system of size $L=4000$ during
$\Delta T=11500$ sampled every 4 timesteps $\delta t=0.16$.
(periodic boundary conditions, transient discarded).
Parts (a) and (c): $\nu=0$, $b_1=1$, $b_3=0.9$.
Parts (b) and (d): $\nu=0$, $b_1=1$, $b_3=0.7$.

\end{itemize}

\end{document}